\documentclass[twocolumn, showkeys, showpacs, preprintnumbers,amsmath,amssymb, prl]{revtex4}

\usepackage{graphicx}
\usepackage{dcolumn}
\usepackage{bm}
\usepackage{epsfig}

\begin{document}

\title{Kink State in a Stack of Intrinsic Josephson Junctions in Layered High-$T_c$ Superconductors and Terahertz Radiation}

\author{Shizeng Lin\(^{1,2}\) and Xiao Hu\(^{1,2,3}\)}

\affiliation{\(^{1}\)WPI Center for Materials
Nanoarchitectonics, National Institute for Materials Science, Tsukuba 305-0044, Japan\\
\(^{2}\)Graduate School of Pure and Applied Sciences, University of
Tsukuba, Tsukuba 305-8571, Japan\\
\(^{3}\)Japan Science and Technology Agency, 4-1-8 Honcho,
Kawaguchi, Saitama 332-0012, Japan}
\date{\today}

\begin{abstract}
A new family of dynamic states are found in a stack of inductively
coupled intrinsic Josephson junctions in the absence of an external
magnetic field. In this state, $(2m_l+1)\pi$ phase kinks with
integers $m_l$'s stack along the c axis and lock neighboring
junction together. Large dc power is pumped into plasma oscillation
via kinks at the cavity resonance. The plasma oscillation is uniform
along the $c$ axis with the frequency satisfying the ac Josephson
relation. Thus this state supports strong terahertz radiation and
seems to be compatible with the recent experimental observations.
\end{abstract}



\maketitle

The discovery of the intrinsic Josephson effect \cite{Kleiner92} in
highly anisotropic layered high-$T_c$ superconductors has opened up
a new direction for generation of terahertz electromagnetic (EM)
waves, which have many promising applications in materials science,
biology, security check and so on\cite{Ferguson02,Tonouchi07}. Much
effort has been taken to investigate the feasibility of such a
technique. One idea is to use motion of Josephson vortices lattice
to excite Josephson plasma, and it attains certain degree of success
\cite{Koyama95,Bulaevskii06,Bae07,szlin08,Rakhmanov09}. In 2007, it
was demonstrated surprisingly that a coherent terahertz EM waves
were emitted from a mesa of $\rm{Bi_2Sr_2CaCu_2O_{8+\delta}}$(BSCCO)
single crystal even in the absence of magnetic field
\cite{Ozyuzer07}. The experiments \cite{Ozyuzer07,kadowaki08} raise
several questions on the mechanism of radiation. First, although the
working principle is certainly the ac Josephson effect, it is not
transparent how dc input power is pumped into ac plasma oscillation,
rather than being dissipated as Joule heating. Secondly, how can
 in-phase plasma oscillation can be attained in all junctions?

The experiments attracted many theoretical
attentions\cite{szlin08b,Koshelev08b,Tachiki09,Koyama09}. A new
dynamic state is proposed to explain the
experiments\cite{szlin08b,Koshelev08b}. In this state, there are
$(2m_l+1)\pi$ kinks localized at the center of the mesa and stacked
periodically along the $c$ axis. The kinks help to pump large energy
into plasma oscillation, and a part of which is emitted outside. The
kink state has already captured the key observations of experiments,
and is believed to be relevant to the emission in experiments.

The inductively coupled sine-Gordon equation appropriately describes the dynamics of gauge invariant phase difference
$P_l$ in a stack of intrinsic Josephson junctions(IJJs)
\begin{equation}\label{eq1}
\partial_x^2 P_l=(1-\zeta \Delta^{(2)})(\sin P_l +\beta \partial_t P_l +\partial_t^2 P_l -J_{\rm{ext}}),
\end{equation}
where $P_l$ is the gauge-invariant phase difference at the $l$th junction, $\Delta^{(2)}Q_l=Q_{l+1}+Q_{l-1}-2Q_{l}$,
$\zeta=\lambda_{ab}^2/(s+D)^2$ the inductive coupling, $\beta$ the normalized conductance, $J_{\rm{ext}}$ the bias
current. $\lambda_{ab}$ is the penetration depth along the $c$ axis and $s$($D$) is the thickness of superconducting
(insulating) layer. Length is normalized by penetration depth in the $ab$ plane $\lambda_{c}$ and time is normalized by
Josephson plasma frequency\cite{szlin08b}. The typical value of $\beta$ is $0.02$ and $\zeta$ is $7.1\times10^4$ for
BSCCO. Here we have neglected the variation of the amplitude of order parameter. The treatment is reasonable because
the Josephson coupling is extremely small in comparison to the condensation energy. The change in $P_l$, which is
associated with that in Josephson energy, should not cause appreciable change in the amplitude, which is associated
with the condensation energy.

In experiments, the thickness of a stack of IJJs is much smaller than the wavelength of electromagnetic wave, which
makes the EM wave transmission from IJJs to outside very difficult. Same situation happens in a thin capacitor. Such a
phenomenon is knows as impedance mismatch. The mismatch in impedance was first formulated explicitly by L. N.
Bulaeviskii and A. E. Koshelev\cite{Bulaevskii06PRL}, and then confirmed by experiments \cite{Kadowaki09a} with a
cylindrical mesa \cite{Hu08,Hu09}. As a good approximation, we may neglect the effect of radiation on the dynamics of
phase. In the present paper, we will use the non-radiating boundary condition $\partial_x P_l=0$. A self-consistent
treatment will be reported elsewhere.

\begin{figure}[t]
\psfig{figure=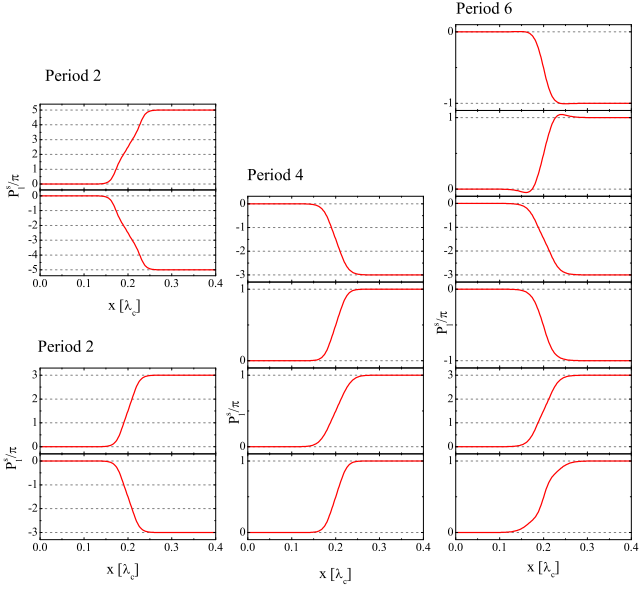,width=\columnwidth} \caption{\label{f1a}(Color online). Several examples of periodic
arrangement of $P_l^s$ in a stack of IJJs calculated from Eq. (\ref{eq3}). Here only even period is shown.}
\end{figure}

Equation (\ref{eq1}) allows for a variety of dynamic states, such as state without kink, state with kink and state with
solitons, each of which has unique electrodynamics properties \cite{szlin09a}. The state with kink\cite{szlin08b} is
described by
\begin{equation}\label{eq2}
P_l(x,t)=\omega t+P_l^s(x) +{\rm{Re}}[-i g(x)\exp(i \omega t)],
\end{equation}
where the firs term at the r.h.s. is the rotating phase with frequency $\omega$, the second term the static kink and
the last term the plasma oscillation whose frequency obeys the ac Josephson relation. The plasma oscillation is uniform
in all junctions which can be realized in a thick stack of IJJs. For the first cavity mode, $g(x)=A_1 \cos k_1 x$ with
$k_j\equiv j\pi/L_x$. It will be shown later that $g(x)$ contains other modes, such as $k_0, k_2, k_3...$ as well, but
it is sufficient to take only $k_1$ mode when $A_1<1$. The solution with kink in Eq. (\ref{eq2}) is not written
artificially. Actually, it is observed repeatedly by computer simulations\cite{szlin08b}, which strongly suggests that
this solution is very stable.

For ease of theoretically treatment, we consider the region with $A_1<1$. Substituting Eq. (\ref{eq2}) into Eq.
(\ref{eq1}) and linearizing the plasma part\cite{szlin09a}, we have the equation for the static kink $P_l^s$
\begin{equation}\label{eq3}
\partial _x^2 P_l^s  =  \frac{{iA_1 \zeta }}{2}\cos (k_1 x)\Delta ^{(2)} \exp ( - iP_l^s ).
\end{equation}
Equation (\ref{eq3}) has $(2m_l+1)\pi$ kink solutions. The operator $\Delta^{(2)}$ allows for a variety of arrangements
of $P_l^s$ along the $c$ axis.

Here we consider several typical periodic arrangements, i.e. $P_l^s=P_{l+\tau}^s$ with $\tau$ the period. The results
for $\tau=2, 4, 6$ are shown in Fig. \ref{f1a}. $(2m_l+1)\pi$ kinks are stacked periodically along the $c$ axis, which
takes the advantage of huge inductive coupling to lock junction together. The kink runs sharply from $0$ to
$(2m_l+1)\pi$ in the region of width $\lambda_P=1/\sqrt{\zeta |A_1|}$. In the BSCCO system, $\zeta\approx10^5$ which
renders the kink almost a step function. The huge inductive coupling also makes the kink very rigid. As shown in Fig.
\ref{f1a}, even for the same period, there are many different ways to pile up the kinks. Thus the kink state occupies
finite volume in the phase space, which makes the state easy to access. The phase kink for higher cavity modes can be
straightforwardly constructed from the fundamental one. The kink is always at the nodes of oscillating electric field.
\begin{figure}[t]
\psfig{figure=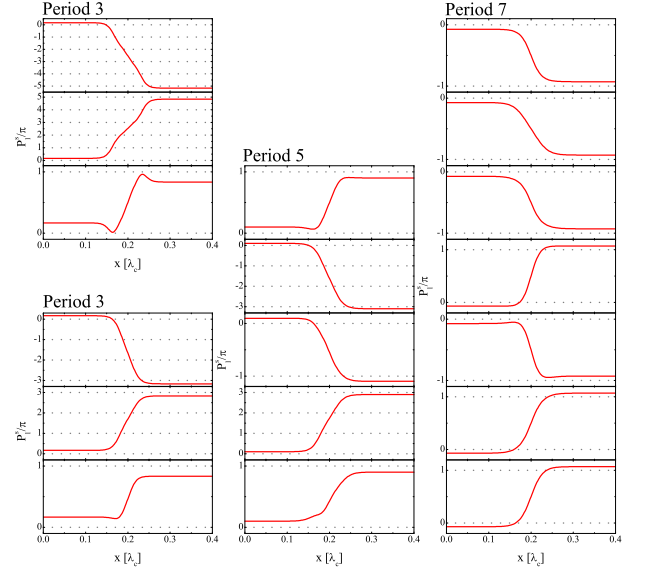,width=\columnwidth} \caption{\label{f1b}(Color online). Several examples of periodic
arrangement of $P_l^s$ in a stack of IJJs calculated from Eq. (\ref{eq3}). Here only odd period is shown.}
\end{figure}

With the relation between magnetic field and phase $(1-\zeta\Delta^{(2)})B_l=\partial_x P_l$, it is easy to show that
flux associated with the static phase kink $P_l^s$ is $(1-\zeta\Delta^{(2)})\Phi_l^s=(1-\zeta\Delta^{(2)})D\int_0^{L_x}
B_l^s dx=D(2 m_l+1)\pi$. Thus the static vortices is quantized, depending on $P_l^s$ and the arrangement of $P_l^s$.

Since no external magnetic field is applied, the total dc flux is
zero in the IJJs, i.e. $\sum_l\Phi_l^s=0$. Thus the value of kink
summed over all junctions vanishes, i.e. $\sum_l(2m_l+1)=0$. For
even period such as $\tau=2, 4, 6$ shown in Fig. \ref{f1a}, the
$(2m_l+1)\pi$ kinks commensurate with the period very well. But for
odd period, there is no way to put $(2m_l+1)\pi$ kinks in junctions
while keeping total dc flux zero, which makes the kinks
incommensurate with IJJs. As depicted in Fig. \ref{f1b}, the kinks
adjust themselves in order to eliminate the static flux. The
resulting kinks deviate slightly from $(2m_l+1)\pi$.

From the equation for the frequency component $\omega t$, we obtain the expression for $A_1$

\begin{equation}\label{eq5}
A_1  = \frac{{F_1 }}{{ik_1^2  - i\omega ^2  - \beta \omega }}
\end{equation}
with
\begin{equation}\label{eq6}
F_1  = \frac{{ - 2i}}{{L_x }}\int\limits_0^{L_x } {\exp (iP_l^s )\cos (k_1 x)dx}
\end{equation}
$F_1$ represents the coupling between the cavity mode and plasma.
Here only the mode $k_1$ of $\exp(iP_l^s)$ is taken. In principle,
there are many different modes $k_j$, and we have to write it down
in $g(x)$ at the very beginning. However only the mode $k_1$ is
dominant when the plasma oscillation is not strong. The static kink
$P_l^s$ plays a role of pumping large dc energy into plasma
oscillation. This becomes more transparent if we look at the total
phase. The total phase is rotating with frequency $\omega$ but it
has kink over the $x$ direction. This type of spatial structure is
compatible with the oscillating electric field, that is the center
of the kink coincides with the nodes of electric field. As the phase
is rotating for the time being, the electric field is excited.

\begin{figure}[t]
\psfig{figure=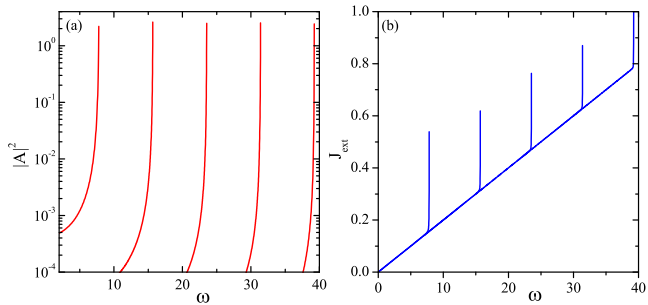,width=\columnwidth} \caption{\label{f2}(Color online). (a) Amplitude of plasma oscillation
$|A|^2$ and (b) \emph{IV} characteristics.}
\end{figure}

Adopting the step-function approximation for $P_l^s$\cite{Hu08}, $A_1$ can be readily evaluated at different $\omega$.
The amplitude at other cavity modes $A_j$ can be easily generalized from Eq. (\ref{eq5}). The results are also depicted
in Fig. \ref{f2}. There is a big enhancement in the amplitude at each cavity resonance, which makes the higher
frequency harmonics visible in the frequency spectrum as calculated by computer simulation\cite{szlin08b}. The higher
frequency harmonics correspond one by one to the high modes in $\exp(iP_l^s)$.

The radiation power can be calculated by using an effective
impedance $Z$ (for simplicity, $Z$ is taken as real). When $Z$ is
large, the radiation can be treated as perturbation, and the
radiation power is give by $S=\omega^2|A_1|^2/2Z$. The dependence of
$S$ on $Z$ is nontrivial. For a large $Z$, much more power can be
pumped into plasma oscillation, and $|A_1|$ is larger than that of
smaller $Z$, which makes the radiation power large. In Ref.
\cite{szlin08b}, by solving Eq. (\ref{eq1}) with radiating boundary
condition numerically, the maximum power at the first cavity mode is
about $10\rm{mW}$ with a mesa of size similar to the
experiments\cite{Ozyuzer07} even for $Z=1000$. The power estimated
by theoretical calculations is still much larger than that in
experiments. There is an optimal value $Z$ which presumes the
maximum radiation power \cite{Nonomura08}.

For theoretical tractability, we have neglected the effect of radiation on the phase dynamics. It does not mean that
the radiation will kill the kink state. Actually, it was found that the kink state survives even for $Z=10$, but the
height of current steps is reduced because it is difficult to excite cavity resonance with small cavity quality
factor\cite{szlin08b}.

The \emph{IV} characteristics is given by the current conservation law
\begin{equation}\label{eq4}
J_{\rm{ext}}=\beta\omega+\langle\sin P_l\rangle_{xt}= \beta \omega  + \frac{{\beta \omega |F_1 |^2 /4}}{{(k_1^2  -
\omega ^2 )^2  + \beta ^2 \omega ^2 }},
\end{equation}
where $\langle \cdot\cdot\cdot \rangle_{xt}$ represents the average over space and time. As considerable amount of dc
power is converted into plasma oscillation at cavity resonances, the \emph{IV} deviates significantly from linear ohmic
behavior and self-induced current steps appear, as in Fig. \ref{f2}. In the absence of irradiation, the input power
$J_{\rm{ext}}\omega$ is consumed by dissipation by normal current $\beta\omega^2$ and by dissipation in plasma
oscillation $\beta\omega|A_1|^2/4$.

In conclusion, $(2m_l+1)\pi$ kink state is formulated in the present paper. In this state, the plasma oscillation is
uniform along the $c$ axis, while the static phase kinks of $(2m_l+1)\pi$ sit periodically along the $c$ axis. The kink
induces a new type of current steps in the \emph{IV} characteristics. This state supports intensive terahertz
radiations. The recent experiments on THz radiations from BSCCO single crystals can be interpreted in terms of this
state in a consistent way.

This work was supported by WPI Initiative on Materials Nanoarchitronics, MEXT, Japan, CREST-JST, Japan and ITSNEM of
CAS.


\begin{thebibliography}{20}
\expandafter\ifx\csname natexlab\endcsname\relax\def\natexlab#1{#1}\fi \expandafter\ifx\csname
bibnamefont\endcsname\relax
  \def\bibnamefont#1{#1}\fi
\expandafter\ifx\csname bibfnamefont\endcsname\relax
  \def\bibfnamefont#1{#1}\fi
\expandafter\ifx\csname citenamefont\endcsname\relax
  \def\citenamefont#1{#1}\fi
\expandafter\ifx\csname url\endcsname\relax
  \def\url#1{\texttt{#1}}\fi
\expandafter\ifx\csname urlprefix\endcsname\relax\def\urlprefix{URL }\fi \providecommand{\bibinfo}[2]{#2}
\providecommand{\eprint}[2][]{\url{#2}}

\bibitem[{\citenamefont{Kleiner et~al.}(1992)\citenamefont{Kleiner, Steinmeyer,
  Kunkel, and M\"{u}ller}}]{Kleiner92}
\bibinfo{author}{\bibfnamefont{R.}~\bibnamefont{Kleiner \emph{et al.}}},
  \bibinfo{journal}{Phys. Rev. Lett.} \textbf{\bibinfo{volume}{68}},
  \bibinfo{pages}{2394} (\bibinfo{year}{1992}).

\bibitem[{\citenamefont{Ferguson and Zhang}(2002)}]{Ferguson02}
\bibinfo{author}{\bibfnamefont{B.}~\bibnamefont{Ferguson}} \bibnamefont{and}
  \bibinfo{author}{\bibfnamefont{X.~C.} \bibnamefont{Zhang}},
  \bibinfo{journal}{Nat. Mater.} \textbf{\bibinfo{volume}{1}},
  \bibinfo{pages}{26} (\bibinfo{year}{2002}).

\bibitem[{\citenamefont{Tonouchi}(2007)}]{Tonouchi07}
\bibinfo{author}{\bibfnamefont{M.}~\bibnamefont{Tonouchi}},
  \bibinfo{journal}{Nat. Photon.} \textbf{\bibinfo{volume}{1}},
  \bibinfo{pages}{97} (\bibinfo{year}{2007}).

\bibitem[{\citenamefont{Koyama and Tachiki}(1995)}]{Koyama95}
\bibinfo{author}{\bibfnamefont{T.}~\bibnamefont{Koyama}} \bibnamefont{and}
  \bibinfo{author}{\bibfnamefont{M.}~\bibnamefont{Tachiki}},
  \bibinfo{journal}{Solid State Commun.} \textbf{\bibinfo{volume}{96}},
  \bibinfo{pages}{367} (\bibinfo{year}{1995}).

\bibitem[{\citenamefont{Bulaevskii and
  Koshelev}(2006{\natexlab{a}})}]{Bulaevskii06}
\bibinfo{author}{\bibfnamefont{L.~N.} \bibnamefont{Bulaevskii}}
  \bibnamefont{and} \bibinfo{author}{\bibfnamefont{A.~E.}
  \bibnamefont{Koshelev}}, \bibinfo{journal}{J. of Supercond. Novel Magn.}
  \textbf{\bibinfo{volume}{19}}, \bibinfo{pages}{349}
  (\bibinfo{year}{2006}{\natexlab{a}}).

\bibitem[{\citenamefont{Bae et~al.}(2007)\citenamefont{Bae, Lee, and
  Choi}}]{Bae07}
\bibinfo{author}{\bibfnamefont{M.~H.} \bibnamefont{Bae \emph{et al.}}},
  \bibinfo{journal}{Phys. Rev. Lett.} \textbf{\bibinfo{volume}{98}},
  \bibinfo{pages}{027002} (\bibinfo{year}{2007}).

\bibitem[{\citenamefont{Lin et~al.}(2008)\citenamefont{Lin, Hu, and
  Tachiki}}]{szlin08}
\bibinfo{author}{\bibfnamefont{S.~Z.} \bibnamefont{Lin \emph{et al.}}},
  \bibinfo{journal}{Phys. Rev.} \textbf{\bibinfo{volume}{B77}},
  \bibinfo{pages}{014507} (\bibinfo{year}{2008}).

\bibitem[{\citenamefont{Rakhmanov et~al.}(2009)\citenamefont{Rakhmanov,
  Savel'ev, and Nori}}]{Rakhmanov09}
\bibinfo{author}{\bibfnamefont{A.~L.} \bibnamefont{Rakhmanov \emph{et al.}}},
  \bibinfo{journal}{Phys. Rev. B} \textbf{\bibinfo{volume}{79}},
  \bibinfo{pages}{184504} (\bibinfo{year}{2009}).

\bibitem[{\citenamefont{Ozyuzer et~al.}(2007)\citenamefont{Ozyuzer, Koshelev,
  Kurter, Gopalsami, Li, Tachiki, Kadowaki, Yamamoto, Minami, Yamaguchi
  et~al.}}]{Ozyuzer07}
\bibinfo{author}{\bibfnamefont{L.}~\bibnamefont{Ozyuzer \emph{et al.}}},
  \bibinfo{journal}{Science}
  \textbf{\bibinfo{volume}{318}}, \bibinfo{pages}{1291} (\bibinfo{year}{2007}).

\bibitem[{\citenamefont{Kadowaki et~al.}(2008)\citenamefont{Kadowaki,
  Yamaguchi, Kawamata, Yamamoto, Minami, Kakeya, Welp, Ozyuzer, Koshelev,
  Kurter et~al.}}]{kadowaki08}
\bibinfo{author}{\bibfnamefont{K.}~\bibnamefont{Kadowaki \emph{et al.}}},
  \bibinfo{journal}{Physca C}
  \textbf{\bibinfo{volume}{468}}, \bibinfo{pages}{634} (\bibinfo{year}{2008}).

\bibitem[{\citenamefont{Lin and Hu}(2008)}]{szlin08b}
\bibinfo{author}{\bibfnamefont{S.~Z.} \bibnamefont{Lin}} \bibnamefont{and}
  \bibinfo{author}{\bibfnamefont{X.}~\bibnamefont{Hu}}, \bibinfo{journal}{Phys.
  Rev. Lett.} \textbf{\bibinfo{volume}{100}}, \bibinfo{pages}{247006}
  (\bibinfo{year}{2008}).

\bibitem[{\citenamefont{Koshelev}(2008)}]{Koshelev08b}
\bibinfo{author}{\bibfnamefont{A.~E.} \bibnamefont{Koshelev}},
  \bibinfo{journal}{Phys. Rev. B} \textbf{\bibinfo{volume}{78}},
  \bibinfo{pages}{174509} (\bibinfo{year}{2008}).

\bibitem[{\citenamefont{Tachiki et~al.}(2009)\citenamefont{Tachiki, Fukuya, and
  Koyama}}]{Tachiki09}
\bibinfo{author}{\bibfnamefont{M.}~\bibnamefont{Tachiki \emph{et al.}}},
  \bibinfo{journal}{Phys. Rev. Lett.} \textbf{\bibinfo{volume}{102}},
  \bibinfo{pages}{127002} (\bibinfo{year}{2009}).

\bibitem[{\citenamefont{Koyama et~al.}(2009)\citenamefont{Koyama, Matsumoto,
  Machida, and Kadowaki}}]{Koyama09}
\bibinfo{author}{\bibfnamefont{T.}~\bibnamefont{Koyama \emph{et al.}}},
  \bibinfo{journal}{Phys. Rev. B} \textbf{\bibinfo{volume}{79}},
  \bibinfo{pages}{104522} (\bibinfo{year}{2009}).

\bibitem[{\citenamefont{Bulaevskii and
  Koshelev}(2006{\natexlab{b}})}]{Bulaevskii06PRL}
\bibinfo{author}{\bibfnamefont{L.~N.} \bibnamefont{Bulaevskii}}
  \bibnamefont{and} \bibinfo{author}{\bibfnamefont{A.~E.}
  \bibnamefont{Koshelev}}, \bibinfo{journal}{Phys. Rev. Lett.}
  \textbf{\bibinfo{volume}{97}}, \bibinfo{pages}{267001}
  (\bibinfo{year}{2006}{\natexlab{b}}).

\bibitem[{\citenamefont{Kadowaki and \emph{et al.}}(2009)}]{Kadowaki09a}
\bibinfo{author}{\bibfnamefont{K.}~\bibnamefont{Kadowaki \emph{et al.}}}, \bibinfo{journal}{preprint}
  (\bibinfo{year}{2009}).

\bibitem[{\citenamefont{Hu and Lin}(2008)}]{Hu08}
\bibinfo{author}{\bibfnamefont{X.}~\bibnamefont{Hu}} \bibnamefont{and}
  \bibinfo{author}{\bibfnamefont{S.~Z.} \bibnamefont{Lin}},
  \bibinfo{journal}{Phys. Rev. B} \textbf{\bibinfo{volume}{78}},
  \bibinfo{pages}{134510} (\bibinfo{year}{2008}).

\bibitem[{\citenamefont{Hu and Lin}(2009)}]{Hu09}
\bibinfo{author}{\bibfnamefont{X.}~\bibnamefont{Hu}} \bibnamefont{and}
  \bibinfo{author}{\bibfnamefont{S.~Z.} \bibnamefont{Lin}},
  \bibinfo{journal}{arXiv.org:0903.2221}  (\bibinfo{year}{2009}).

\bibitem[{\citenamefont{Lin and Hu}(2009)}]{szlin09a}
\bibinfo{author}{\bibfnamefont{S.~Z.} \bibnamefont{Lin}} \bibnamefont{and}
  \bibinfo{author}{\bibfnamefont{X.}~\bibnamefont{Hu}}, \bibinfo{journal}{Phys.
  Rev. B} \textbf{\bibinfo{volume}{79}}, \bibinfo{pages}{104507}
  (\bibinfo{year}{2009}).

\bibitem[{\citenamefont{Nonomura}(2008)}]{Nonomura08}
\bibinfo{author}{\bibfnamefont{Y.}~\bibnamefont{Nonomura}},
  \bibinfo{journal}{arXiv:0810.3756}  (\bibinfo{year}{2008}).

\end{thebibliography}

\end{document}